\documentstyle[11pt,gh2001-asp,twoside,epsf]{article}
\ifx\epsfannounce\undefined \def\epsfannounce{\immediate\write16}\fi
 \epsfannounce{This is `epsf.tex' v2.7k <10 July 1997>}%
\newread\epsffilein    % file to \read
\newif\ifepsfatend     % need to scan to LAST %%BoundingBox comment?
\newif\ifepsfbbfound   % success?
\newif\ifepsfdraft     % use draft mode?
\newif\ifepsffileok    % continue looking for the bounding box?
\newif\ifepsfframe     % frame the bounding box?
\newif\ifepsfshow      % show PostScript file, or just bounding box?
\epsfshowtrue          % default is to display PostScript file
\newif\ifepsfshowfilename % show the file name if \epsfshowfalse specified?
\newif\ifepsfverbose   % report what you're making?
\newdimen\epsfframemargin % margin between box and frame
\newdimen\epsfframethickness % thickness of frame rules
\newdimen\epsfrsize    % vertical size before scaling
\newdimen\epsftmp      % register for arithmetic manipulation
\newdimen\epsftsize    % horizontal size before scaling
\newdimen\epsfxsize    % horizontal size after scaling
\newdimen\epsfysize    % vertical size after scaling
\newdimen\pspoints     % conversion factor
\def\epsfbox#1{\global\def\epsfllx{72}\global\def\epsflly{72}%
   \global\def\epsfurx{540}\global\def\epsfury{720}%
   \def\lbracket{[}\def\testit{#1}\ifx\testit\lbracket
   \let\next=\epsfgetlitbb\else\let\next=\epsfnormal\fi\next{#1}}%
\def\epsfgetlitbb#1#2 #3 #4 #5]#6{%
   \epsfgrab #2 #3 #4 #5 .\\%
   \epsfsetsize
   \epsfstatus{#6}%
   \epsfsetgraph{#6}%
}%
\def\epsfnormal#1{%
    \epsfgetbb{#1}%
    \epsfsetgraph{#1}%
}%
\newhelp\epsfnoopenhelp{The PostScript image file must be findable by
TeX, i.e., somewhere in the TEXINPUTS (or equivalent) path.}%
\def\epsfgetbb#1{%
    \openin\epsffilein=#1
    \ifeof\epsffilein
        \errhelp = \epsfnoopenhelp
        \errmessage{Could not open file #1, ignoring it}%
    \else                       %process the file
        {%                      %start a group to contain catcode changes
            \chardef\other=12
            \def\do##1{\catcode`##1=\other}%
            \dospecials
            \catcode`\ =10
            \epsffileoktrue         %true while we are looping
            \epsfatendfalse     %[02-Jul-1996]: add forgotten initialization
            \loop               %reading lines from the EPS file
                \read\epsffilein to \epsffileline
                \ifeof\epsffilein %then no more input
                \epsffileokfalse %so set completion flag
            \else                %otherwise process one line
                \expandafter\epsfaux\epsffileline:. \\%
            \fi
            \ifepsffileok
            \repeat
            \ifepsfbbfound
            \else
                \ifepsfverbose
                    \immediate\write16{No BoundingBox comment found in %
                                    file #1; using defaults}%
                \fi
            \fi
        }%                      %end catcode changes
        \closein\epsffilein
    \fi                         %end of file processing
    \epsfsetsize                %compute size parameters
    \epsfstatus{#1}%
}%
\def\epsfclipoff{\def\epsfclipstring{\ifepsfdraft\space clip\fi}}%
\epsfclipoff % default for dvips is OFF
\def\epsfspecial#1{%
     \epsftmp=10\epsfxsize
     \divide\epsftmp\pspoints
     \ifnum\epsfrsize=0\relax
       \includegraphics{\ifepsfdraft}%
     \else
       \epsfrsize=10\epsfysize
       \divide\epsfrsize\pspoints
       \includegraphics{\ifepsfdraft}%
     \fi
}%
\def\epsfframe#1%
{%
  \leavevmode                   % so we can put this inside
  \setbox0 = \hbox{#1}%
  \dimen0 = \wd0                                % natural width of argument
  \advance \dimen0 by 2\epsfframemargin         % plus width of 2 margins
  \advance \dimen0 by 2\epsfframethickness      % plus width of 2 rule lines
  \vbox
  {%
    \hrule height \epsfframethickness depth 0pt
    \hbox to \dimen0
    {%
      \hss
      \vrule width \epsfframethickness
      \kern \epsfframemargin
      \vbox {\kern \epsfframemargin \box0 \kern \epsfframemargin }%
      \kern \epsfframemargin
      \vrule width \epsfframethickness
      \hss
    }% end hbox
    \hrule height 0pt depth \epsfframethickness
  }% end vbox
}%
\def\epsfsetgraph#1%
{%
   \leavevmode
   \hbox{% so we can put this in \begin{center}...\end{center}
     \ifepsfframe\expandafter\epsfframe\fi
     {\vbox to\epsfysize
     {%
        \ifepsfshow
            \vfil
            \hbox to \epsfxsize{\epsfspecial{#1}\hfil}%
        \else
            \vfil
            \hbox to\epsfxsize{%
               \hss
               \ifepsfshowfilename
               {%
                  \epsfframemargin=3pt % local change of margin
                  \epsfframe{{\tt #1}}%
               }%
               \fi
               \hss
            }%
            \vfil
        \fi
     }%
   }}%
   \global\epsfxsize=0pt
   \global\epsfysize=0pt
}%
\def\epsfsetsize
{%
   \epsfrsize=\epsfury\pspoints
   \advance\epsfrsize by-\epsflly\pspoints
   \epsftsize=\epsfurx\pspoints
   \advance\epsftsize by-\epsfllx\pspoints
   \epsfxsize=\epsfsize{\epsftsize}{\epsfrsize}%
   \ifnum \epsfxsize=0
      \ifnum \epsfysize=0
        \epsfxsize=\epsftsize
        \epsfysize=\epsfrsize
        \epsfrsize=0pt
      \else
        \epsftmp=\epsftsize \divide\epsftmp\epsfrsize
        \epsfxsize=\epsfysize \multiply\epsfxsize\epsftmp
        \multiply\epsftmp\epsfrsize \advance\epsftsize-\epsftmp
        \epsftmp=\epsfysize
        \loop \advance\epsftsize\epsftsize \divide\epsftmp 2
        \ifnum \epsftmp>0
           \ifnum \epsftsize<\epsfrsize
           \else
              \advance\epsftsize-\epsfrsize \advance\epsfxsize\epsftmp
           \fi
        \repeat
        \epsfrsize=0pt
      \fi
   \else
     \ifnum \epsfysize=0
       \epsftmp=\epsfrsize \divide\epsftmp\epsftsize
       \epsfysize=\epsfxsize \multiply\epsfysize\epsftmp
       \multiply\epsftmp\epsftsize \advance\epsfrsize-\epsftmp
       \epsftmp=\epsfxsize
       \loop \advance\epsfrsize\epsfrsize \divide\epsftmp 2
       \ifnum \epsftmp>0
          \ifnum \epsfrsize<\epsftsize
          \else
             \advance\epsfrsize-\epsftsize \advance\epsfysize\epsftmp
          \fi
       \repeat
       \epsfrsize=0pt
     \else
       \epsfrsize=\epsfysize
     \fi
   \fi
}%
\def\epsfstatus#1{% arg = filename
   \ifepsfverbose
     \immediate\write16{#1: BoundingBox:
                  llx = \epsfllx\space lly = \epsflly\space
                  urx = \epsfurx\space ury = \epsfury\space}%
     \immediate\write16{#1: scaled width = \the\epsfxsize\space
                  scaled height = \the\epsfysize}%
   \fi
}%
{\catcode`\%=12 \global\let\epsfpercent=%\global\def\epsfbblit{%BoundingBox}}%
\global\def\epsfatend{(atend)}%
\long\def\epsfaux#1#2:#3\\%
{%
   \def\testit{#2}%             % save second character up to just before colon
   \ifx#1\epsfpercent           % then first char is percent (quick test)
       \ifx\testit\epsfbblit    % then (slow test) we have %%BoundingBox
            \epsfgrab #3 . . . \\%
            \ifx\epsfllx\epsfatend % then ignore %%BoundingBox: (atend)
                \global\epsfatendtrue
            \else               % else found %%BoundingBox: llx lly urx ury
                \ifepsfatend    % then keep parsing ALL %%BoundingBox lines
                \else           % else stop after first one parsed
                    \epsffileokfalse
                \fi
                \global\epsfbbfoundtrue
            \fi
       \fi
   \fi
}%
\def\epsfempty{}%
\def\epsfgrab #1 #2 #3 #4 #5\\{%
   \global\def\epsfllx{#1}\ifx\epsfllx\epsfempty
      \epsfgrab #2 #3 #4 #5 .\\\else
   \global\def\epsflly{#2}%
   \global\def\epsfurx{#3}\global\def\epsfury{#4}\fi
}%
\def\epsfsize#1#2{\epsfxsize}%
\setcounter{page}{141}

\def\edcomment#1{\iffalse\marginpar{\raggedright\sl#1\/}\else\relax\fi}
\marginparwidth 1.25in
\marginparsep .125in
\marginparpush .25in
\reversemarginpar

\begin{document}
\title{Formation and evolution of bars in disc galaxies}
 \author{E. Athanassoula}
\affil{Observatoire de Marseille, 2 Place Le Verrier, 13248 Marseille
 cedex 04, France}

\begin{abstract}
I follow a bar from its formation, via its evolution, to its destruction
and, perhaps, regeneration. I discuss the main features at each stage
and particularly the role of the halo. Bars can form even in
sub-maximum discs. In fact, such bars can be stronger than bars which
have grown in maximum discs. This is due to the response of the halo
and, in particular, to the exchange of energy and angular momentum
between the disc particles constituting the bar and the halo particles
at resonance with it. The bar slowdown depends on the initial central
concentration of the halo and the initial value of the disc
$Q$. Contrary to the halo mass distribution, the disc changes its
radial density profile considerably during the evolution. Applying the
Sackett criterion, I thus find that discs become maximum in many
simulations in which they have started off as sub-maximum. I briefly discuss
the evolution if a gaseous component is present, as well as the destruction and
regeneration of bars.
\end{abstract}

\section{Introduction}

Bars can form spontaneously in galactic discs (e.g. Miller,
Prendergast \& Quirk 1970, Hohl 1971). They then evolve over a
large number of rotations, changing their length, strength, shape and
angular frequency. During this period they can
drive spiral or ring formation, push gas to the center-most parts of
the galaxy and thus
trigger starbursts and activity in the nucleus, and interact with the
outer disc, bulge and/or halo by exchanging energy and angular momentum with
them. I will discuss a few of these processes, relying -- due to the
strongly nonlinear nature of bars -- mainly on
results of $N$-body simulations. 

\section{Simulations}
\label{sec:simulations}

The results presented in sections 3 to 6  come from a preliminary 
analysis of well over 100 simulations run on our GRAPE-5 systems (Kawai et al.
2000). Six of these have been discussed by Athanassoula \& Misiriotis
(2002, hereafter AM)
and Athanassoula (2002, hereafter A02), where more information on the numerical
techniques can be found. Initially the disc is always exponential
radially, and follows an $sech^2$ law with scalelength $z_0$ in the
vertical direction. Its initial $Q$ value (hereafter $Q_{init}$) does
not vary with radius.   
The disc mass ($M_d$) and initial scalelength ($h$) will be used as
units of mass and length in the simulations. The halo is initially
isotropic and non-rotating, and follows the density law 

\begin{equation}
\rho_h (r) = \frac {M_h}{2\pi^{3/2}}~~ \frac{\alpha}{r_c} ~~\frac {exp~(-r^2/r_c^
2)}{r^2+\gamma^2},
\end{equation}

\noindent
where $\alpha$ is a normalisation constant, $M_h$ is the mass of the halo
and $\gamma$ and $r_c$ are scalelengths. This functional forms
contains three free parameters and thus allows a large
flexibility. For relatively large values of $\gamma$ the halo profile
resembles those used by observers in rotation curve decompositions,
and is thus observationally motivated. On the contrary, for
sufficiently small values of $\gamma$ the halo rotation curve rises
very steeply, thus mimicking a cusp in the halo distribution. Indeed,
the core can be chosen sufficiently small to be hidden by the
softening length. In all the simulations discussed here, except for
the ones mentioned in the end of section 6,
$M_h$ = 5 and $r_c$ = 10. 

The total number of particles per simulation varies between 1 and 
1.5 million. I adopted a softening of 0.0625 and a time-step
of 0.015625. I assessed 
the numerical robustness of my results by trying in a few cases
double the number of particles, different values of the softening
and time step, as well as direct summation (with half the number of
particles) and a non-GRAPE tree code. Various calibrations are
possible. AM proposed to set the unit of length equal to 
3.5 kpc and the unit of mass to 5 $\times$ $10^{10}$ $M_{\odot}$.
In this case $t$ = 500 corresponds to 7 Gyrs. 
Alternatively, one could use a considerably smaller value in kpc for
the disc scalelength, arguing that this is the {\it initial} disc
scalelength and that this length will increase with time. Thus opting
for a 2 kpc scalelength would set $t$ = 500 to roughly 3 Gyrs. Time, in
particular, may be difficult to calibrate, since numerical parameters,
like the softening and the number of particles, can influence the rate
at which the bar grows in the initial stages of the simulation. Thus
the clock may tick differently, 
at least for certain parts of the evolution, in simulations than in
real galaxies. For these reasons, here
we will stay with computer units, allowing the
reader the freedom to convert them to astronomical units according to
his/her preferences and needs.

The two parameters in the initial conditions influencing most the
evolution are the central  
concentration of the halo and $Q_{init}$. The former is given by the parameter 
$\gamma$, which can be thought of as the core radius of the
halo. 

\section{The effect of halo central concentration}
\label{sec:haloconcen}

\begin{figure} 
\plotone{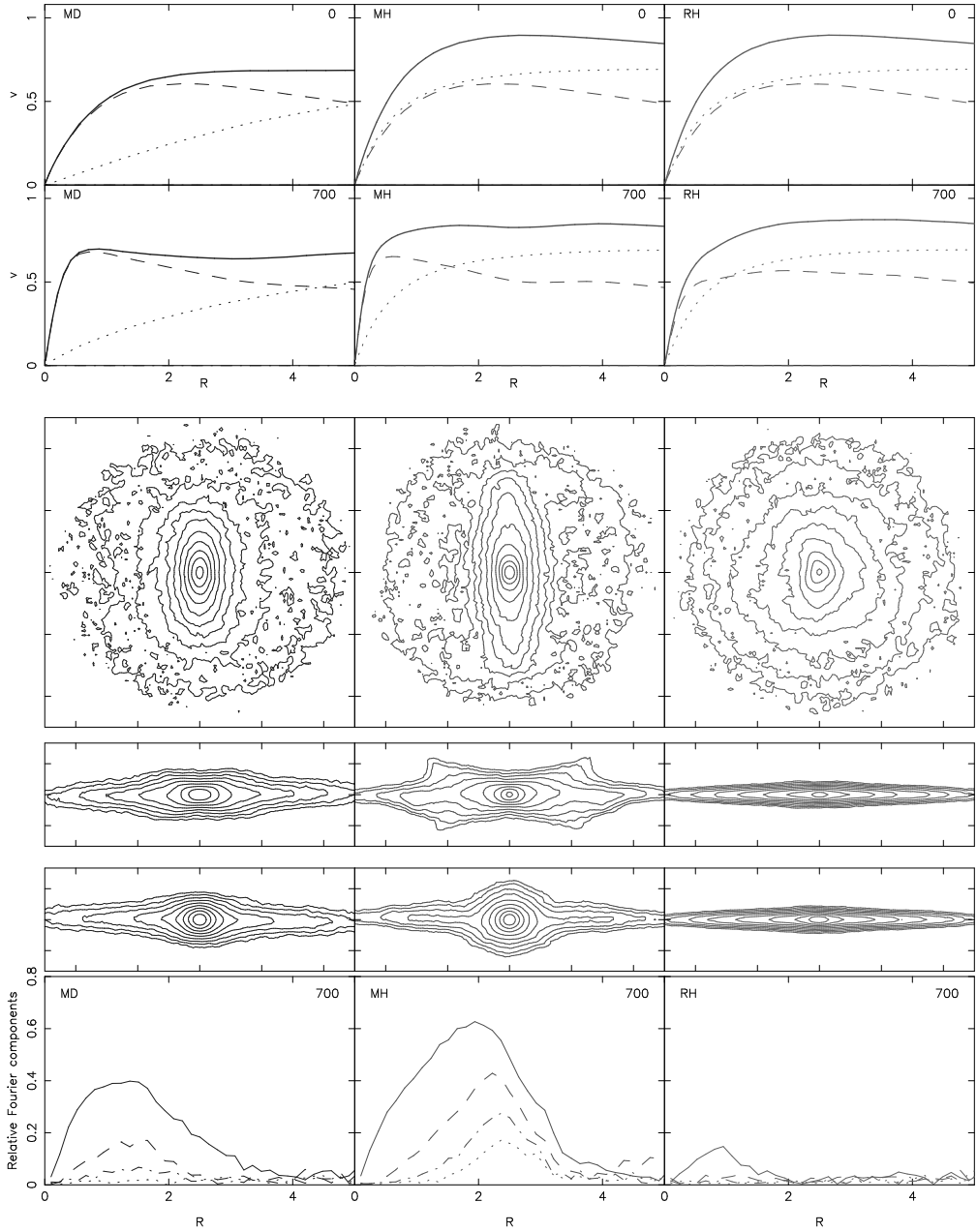}
\caption{The two upper rows of panels give the circular
    velocity curves (solid lines) and the contribution of the disc
    (dashed) and the halo (dotted). The first row
    is for $t$ = 0, the second for $t$ = 700. The next three
    rows give the isodensities of the disc component when seen
    face-on (third row) and edge-on, again for
    $t$ = 700. In the
    fourth row the bar is seen side-on and in the fifth end-on. 
    The size of the square box 
    for the third row of panels is 10$h$. The
last row shows the relative Fourier 
components as a function of radius (AM). The $m$ = 2, 4, 6 and 8
components are given by solid, dashed, dot-dashed and dotted lines.
} 
\label{fig:basic}
\end{figure}

Figure~\ref{fig:basic}\footnote{In the third row of panels, giving the
face-on view, I used density values for the isocontours such as
to show the structure in the bar region, but leaving out the disc
region at larger radii.} compares three simulations. The one 
illustrated in the 
left column of panels has $\gamma$ = 5 and its inner parts are disc 
dominated from the onset (upper left panel). 
I will hereafter refer to it as model MD, for 
massive disc. The middle and right panels refer to simulations with 
$\gamma$ = 0.5. Here the halo is much more centrally concentrated and 
contributes to the total circular velocity curve somewhat more than 
the disc up to roughly two disc scalelengths, and considerably 
more at larger radii (middle and left upper panels).
I will refer to these two simulations as MH, for massive halo, and RH,
for rigid halo, respectively. Indeed in simulation MH, as in MD, the 
halo is live, i.e. composed of particles, while in simulation RH it 
is rigid, i.e. is described by a force and potential imposed on the 
disc particles. All three simulations start off with $Q_{init}$ = 1.2.

The differences between the results of simulations MD and MH are very
important. The bar in model MH is longer and stronger than that in model MD. 
Its isodensities are more rectangular-like and, seen edge-on with the
bar seen side-on, it displays an `X' shape, while for the same
orientation MD shows a boxy structure. Both shapes have been observed
in edge-on galaxies. Seen end-on, MD again shows a
boxy structure, while MH shows a spheroidal feature of considerable
vertical extent. If such a feature was observed in an edge-on galaxy,
it would be erroneously classified as a sizeable bulge component. Thus
a number of observed bulges could in fact be bars seen end-on. A
final, very important difference can be seen by comparing the relative
Fourier components of the two models (AM). Model MH has a considerably
larger $m$ = 2 component. More important, it has also considerable $m$ =
4, 6 and 8 components, very much in agreement with what is observed in
early type barred spirals (e.g. Ohta 1996). On the contrary the Fourier
components of model MD are more reminiscent of those of late type
galaxies. All these differences are surprising and most of them are
contrary to previous wisdom, since {\it the stronger bar is found 
in the more halo dominated galaxy.} 

\section{Evolution of some basic quantities}
\label{sec:evol}

The evolution does not stop after the initial phases of bar formation.
I will here describe the evolution of some basic properties, namely the 
strength and pattern speed of the bar and the angular momentum and mass 
distribution in the disc and halo components.

\subsection{Pattern speed}

\begin{figure} 
\plottwo{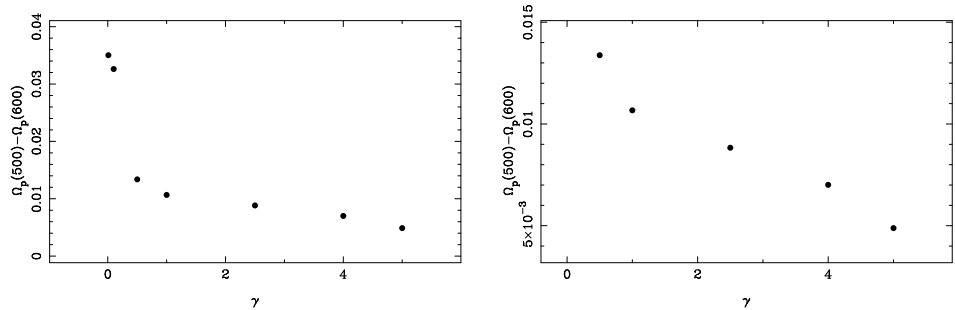}{atha-fig2b.ps}
\caption{Left panel : Slowdown of the bar pattern speed between times
500 and 600 as a function of the halo central concentration. For these
simulations initially $Q_{init}$ = 1.2 and $z_0$ = 0.2. Right panel : Blow-up
of the lower part of the left panel, showing 
that the trend with central concentration is clear even for higher
values of $\gamma$. 
 }
\label{fig:omp_gamma}
\end{figure}

Due to the interaction with the halo and the outer disc, the bar 
gradually slows down (see e.g. Tremaine \& Weinberg 1984,
Weinberg 1985, Combes et al. 1990, Little \& Carlberg 1991, Hernquist
\& Weinberg 1992, Athanassoula 1996, Debattista \& Sellwood 1998 and
2000). The rate at which this happens depends crucially on the two
parameters which mainly influence the evolution, namely $\gamma$ and
$Q_{init}$.  

Simulations with more centrally concentrated haloes, i.e. of 
MH-type, show a considerably larger slowdown rate than simulations
of MD-type. This is clearly seen in Fig.~\ref{fig:omp_gamma}, where I
plot the slowdown of the bar between times 500 and 600 as a function
of $\gamma$. In particular for the smallest values of $\gamma$,
i.e. 0.01 and 0.1, the slowdown is non negligible. Even for
larger values, however, there is a definite trend of decreasing slowdown rate
with increasing $\gamma$ (right panel). This is in good qualitative agreement 
with the results of Debattista   
\&  Sellwood (1998, 2000), who showed that in cases with massive 
discs there is an acceptable slowdown of the bar, contrary to 
simulations with massive halo components. A quantitative comparison
will be given elsewhere.

\begin{figure} 
\plottwo{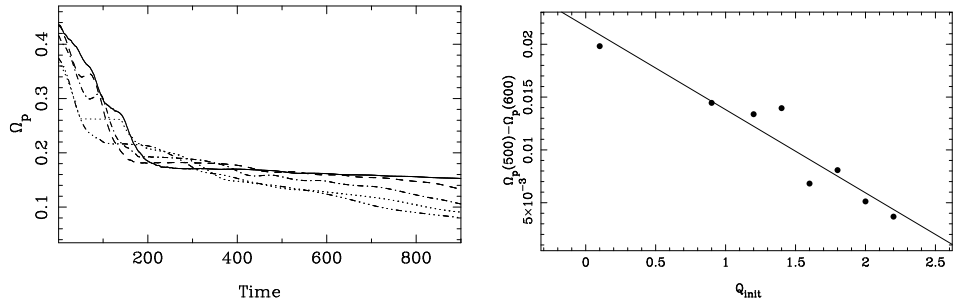}{atha-fig3b.ps}
\caption{Left panel : Bar pattern speed as a function of time, for
$Q_{init}$ = 1.4 (dot-dot-dot-dashed), 1.6 (dotted), 1.8
(dot-dashed), 2 (dashed) and 2.2
(solid line). Right panel : Slowdown of the bar between times
500 and 600 as a function of $Q_{init}$. For these simulations
initially $\gamma$ = 0.5 and $z_0$ = 0.2.  
}
\label{fig:Q}
\end{figure}

The halo mass distribution, however, is not the only parameter 
influencing the slowdown of the bar. The left panel of Fig.~\ref{fig:Q} shows 
the evolution of the pattern speed with time for simulations
which have initially identical live haloes and identical mass distributions
in the disc component, but different $Q_{init}$. It is clear that
$Q_{init}$ influences strongly the bar slowdown, in
the sense that bars in initially colder discs show a much stronger
slowdown rate than bars in initially hotter discs. This is further
explicited in the right panel, where I give the slowdown of the bar
between times 500 and 600 as a function of  
$Q_{init}$. The solid line is simply a least squares fit to the
data to guide the eye and does not imply that the decrease should be linear. 

Thus the slowdown rate is a function of at least two parameters, $Q_{init}$
and $\gamma$, and a particular value of this rate can be achieved not
only for a single ($Q_{init}$, $\gamma$) pair, but for a sequence of
such pairs. 
One should, therefore, refrain from using this rate to 
obtain limits on the values of one of the two parameters, unless the
value of the other is relatively well constrained. This should be kept in
mind when using arguments based on the bar slowdown in the ongoing
debate about whether discs are maximum or sub-maximum, since little is
known about $Q_{init}$. One could argue
that discs were initially very cold, as they were composed mainly of
gas, which formed stars with low velocity dispersion.
On the other hand, it could also be argued that discs had initially
very inhomogeneous mass distributions with 
big lumps of matter, whose interactions could heat the initial disc to
relatively high temperatures. This, together with the inflow of fresh
cool gas from the halo, should then determine $Q_{init}$. 

If the $Q_{init}$ is sufficiently large, then the bar will not
slow down substantially. Furthermore, if $Q_{init}$ is
intermediate, then the increase in bar length (see
\S4.2) will compensate for the slowdown, in the
sense that $r_L / a$ (where $r_L$ the corotation radius and $a$ the
bar length) will remain roughly constant during the evolution. This
will be discussed in 
detail elsewhere (Athanassoula 2002, in preparation). Here let us just
note that, since our present knowledge does not allow us to
have a clear view of the situation, we should refrain from drawing
conclusions, if these are sensitive on the value of $Q_{init}$. Finally, the
slowdown rate could also be dependent on the resolution of the
simulations. However, comparisons of simulations with different
resolutions should necessarily involve only simulations with the same
or similar $Q_{init}$, if these are to give meaningful information
on the effect of resolution. This issue will be further discussed elsewhere
(Athanassoula 2002, in preparation). 

\subsection{Bar strength and length}
\label{subsec:strength}

The strength of the bar and its evolution with time also depend 
crucially on $\gamma$ and $Q_{init}$. The left panel of
Fig.~\ref{fig:sackett}
compares the evolution of the bar strength for the three simulations 
already discussed in section \ref{sec:haloconcen} The bar strength is
here defined as the total $m$ = 2 component of the mass,
i.e. integrated over the surface of the disc, divided by the
corresponding total $m$
= 0 component. Note that in the initial phases of the evolution 
the bar grows faster in simulation MD than in MH, in good agreement with the 
results of the 2D simulations by Athanassoula \& Sellwood (1986). At 
later stages, however, the bar continues growing slowly but steadily in 
simulation MH, compared to hardly, if at all, in simulation MD.
Thus, eventually, the bar in MH becomes considerably stronger 
than in MD. 

The length of the bar in model MD does not show much evolution with
time. On the contrary, for model MH it grows considerably with time.

\subsection{Angular momentum and mass distribution }

During the evolution, the disc gives angular momentum to the
halo (see also Debattista \&  Sellwood 2000) and, as a result, the latter
obtains a net rotation in the same sense as the disc. This effect is stronger
for MH-type models than for MD-type ones. 

As can be seen by comparing the two upper panels of
Fig.~\ref{fig:basic}, the disc mass radial profile evolves considerably, 
becoming much more centrally concentrated. This effect is stronger for
initially colder discs. On the other hand the mass distribution in the
halo component shows little evolution with time (Fig.~\ref{fig:basic}).

\section{Maximum versus sub-maximum discs}
\label{sec:maxdisc}

Observed rotation curves and velocity fields provide useful
information on the gravitational forces that act in galactic discs. It is,
however, not clear what fraction of this force is due to the disc and
the halo components, respectively. This has been the subject of
animated debates, some pieces of evidence arguing that discs are
maximum, i.e. that they dominate in the inner regions, and others that 
their contribution is substantially lower (e.g. e.g. Athanassoula,  
Bosma \& Papaioannou 1987; Bosma 1999, 2000 and these proceedings;
Bottema 1993; Courteau \& Rix 1999;
Kranz, Slyz \& Rix 2001; Sellwood 1999; Weiner, Sellwood \& Williams
2001). 

Sackett (1997) and Bosma (2000) give a simple, straightforward
criterion, allowing us to distinguish maximum from sub-maximum
discs. Consider the ratio
$S = V_{d,max} / V_{tot}$, where $V_{d,max}$ is the circular 
velocity due to the disc component 
and $V_{tot}$ is the total circular velocity, both calculated at a radius 
equal to 2.2 disc scalelengths. According to Sackett (1997) this
ratio has to be at least 0.75 for the disc to be considered
maximum. Of course in the case of strongly  
barred galaxies the velocity field is non-axisymmetric and one should
consider azimuthally averaged rotation curves, or ``circular
velocity'' curves. Furthermore, in the case of strongly 
barred galaxies it is not easy to define a disc scalelength, so it is
better to calculate $S$ at the radius at which the disc  
rotation curve is maximum, which is a well defined radius and 
is roughly equal to 2.2 disc scalelengths in the case of an 
axisymmetric exponential disc. After this small adjustment, we can apply the
above criterion to our simulations. 

\begin{figure} 
\plottwo{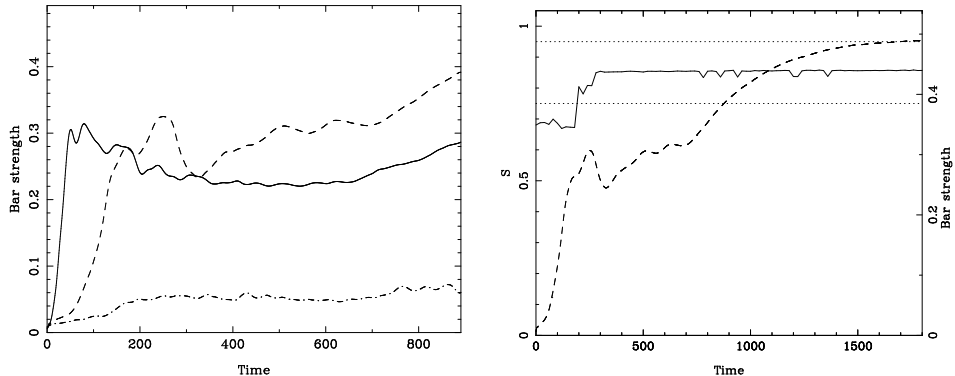}{atha-fig4b.ps}
\caption{Left panel : Bar strength as a function of time for
  simulation MD (solid line), MH (dashed) and RH (dot-dashed). Right
  panel : Evolution of $S$ as a function of time (solid line) for
  simulation MH. 
The two horizontal dotted lines give the limits set by
Sackett for a maximum disc. The dashed line gives a measure of the
strength of the bar.}
\label{fig:sackett}
\end{figure}

As can be seen from Fig.~\ref{fig:basic}, the MD disc 
starts off as maximum, and its importance 
increases further with time. On the other hand simulation
MH starts as sub-maximum, with a value of $S$ not far from the value of 
0.63 advocated by Bottema (1993). The right panel of
Fig.~\ref{fig:sackett} plots the  
evolution of $S$ with time and shows clearly that after roughly
time 200 the disc becomes maximum, and then stays so, at a roughly constant
value of $S$. It is interesting to
note that the rather abrupt jump of the disc from sub-maximum to
maximum happens right after the abrupt increase of the bar strength,
i.e. right after bar formation.

In the above example the disc started as sub-maximum and ended as
maximum, due to the rearrangement of the disc material. Yet
this is not always the case. In cases with strong bulges, small
$\gamma$ and/or large $Q_{init}$, the value of $S$ may not rise sufficiently
for the disc to become maximum. The limiting
value of $\gamma$ for which the disc will become maximum after bar formation 
depends on the values of the remaining parameters of the
simulation. Interpolating from a set of simulations with $Q_{init}$ = 1.2,
$z_0$ = 0.2 and $M_h$ = 5, we find the limiting value of $\gamma$ to
be around 0.4. On the other hand, for initially very cold discs, with $Q_{init}$ = 
0.1, the limiting value of $\gamma$ is much smaller, of the order of
0.13. This last value corresponds to a very fast rising rotation
curve, reminiscent of those obtained with a cuspy halo density profile. 
In fact, since this value of $\gamma$ is only slightly larger than twice the softening 
length, a smaller distance would not be well resolved with the 
present simulations. 

We can thus conclude that whether a disc ends as maximum or as sub-maximum
depends on the central concentration of the halo and on the
initial velocity dispersion in the disc. For highly centrally
concentrated but non-cuspy haloes ($\gamma$ = 0.5), as those favoured
by observers for early type galaxies, an initially sub-maximum disc
will become maximum after the bar has formed. This holds for all
such initial conditions that I tried, except if $Q_{init}$ is
larger than 1.7. On the other hand, if the haloes are cuspy, then the
threshold $Q_{init}$ is considerably smaller. It should also be noted
that the exact position of this threshold may well depend on the
initial $z_0$ and on the halo-to-disc mass ratio
-- which here has been taken to be equal to 5 -- and perhaps also on
numerical parameters like the softening. Thus, although the above
discussion argues that it is quite likely that discs of strongly
barred galaxies are maximum, a considerable amount of work is still
necessary in order to produce a rigid criterion. 

\section{The role of the halo}
\label{sec:halo}

The strongest differences in Fig.~\ref{fig:basic} are between the
middle (model MH) and right (model RH) columns of panels. Indeed, the
face-on view of model 
MH shows a very strong bar and the edge-on one a very strong peanut,
or rather an `X'-type feature. On the contrary, model RH shows at most
a slight oval distortion in the inner parts of the face-on view and
not even a boxiness in the edge-on views. Similarly, the rotation
curves in the second row of panels reveal that the disc material for
model MH has 
considerably concentrated towards the center, while there is only a very
small such effect for RH. Since the two models are initially
identical in everything, except that the halo of RH is rigid, while that
of MH is live, it must be the halo response that is responsible
for the big differences.  

\begin{figure} 
\plotfiddle{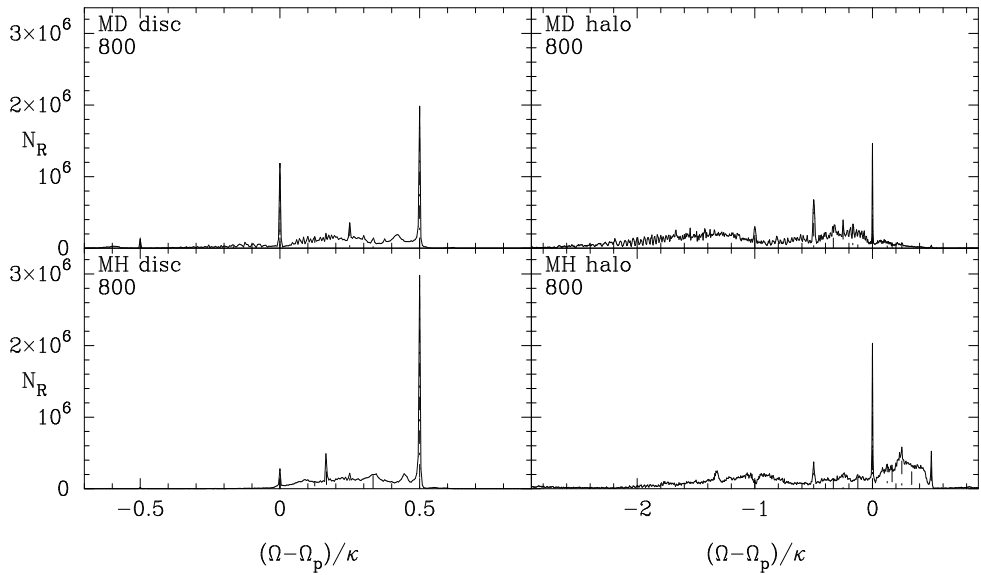}{6.7cm}{0}{75}{80}{-220}{-200}
\caption{Number density, $N_R$,  of particles as a
  function of the frequency ratio $R_f = (\Omega -
\Omega_p) / \kappa$. The results for the halo
  component have been rescaled, so as to take into account the different
  number of particles in the disc and halo components, and thus allow
  immediate comparisons.}
\label{fig:reson}
\end{figure}

In order to understand the halo response we need first to understand
its orbital structure. We thus need to calculate, at a given
time $t$, the principal frequencies of the halo and disc orbits,
i.e. the values of the angular frequency $\Omega$, the epicyclic
frequency $\kappa$ and the vertical frequency $\kappa_z$ for
each orbit (A02). For this, I calculate the total potential and the bar
pattern speed at this time. In this potential, now considered
as stationary in a frame of reference rotating with the
measured pattern speed, I calculate the orbits of 100~000 disc and 100~000 halo
particles and from these their basic frequencies. 
This reveales that a considerable number of
halo stars are in resonance with the bar. Fig.~\ref{fig:reson} plots
the number of particles (orbits), 
$N_R$, that have a frequency ratio $R_f = (\Omega - \Omega_p) / \kappa$
within a bin of a given width centered on a value of this ratio, 
plotted as a function of $R_f$. The distribution is far from uniform
and shows high peaks at the main resonances, both for the disc
and the halo components. Disc particles at inner Lindblad
resonance will emit energy 
and angular momentum (Lynden-Bell \& Kalnajs 1972) which can be
absorbed either by disc particles at corotation and outer Lindblad
resonance, or by halo resonant
particles. Since the bar is a negative angular momentum perturbation,
the halo, by taking angular momentum from it, helps it grow. Thus
haloes, instead of stabilising discs, as was previously thought, can
help bars grow. This new instability mechanism (A02) can account for
strong bars in cases with strong halo components, and explains
why the bar in model MH is stronger than that in model MD. 

However, one should not extrapolate this behaviour further to ever
stronger halo components and conclude that the stronger the halo the
stronger the bar should be. There will obviously be a point where the
disc will be comparatively too light to emit all the angular momentum
the halo can absorb.
For simulations with $M_h / M_d$ = 5 the discs proved to be bar
unstable for all the values of $\gamma$ I 
tried. This, however, is not always the case for smaller disc-to-halo mass
ratios. Thus no bar formed for $M_h / M_d$ = 10 and $\gamma$ =
0.5 at least up to $t$ = 900, and presumably much longer. This shows
that, although bars can grow in sub-maximum discs, they can not grow
in galaxies with an arbitrarily small disc contribution. 

\section{Evolution in the presence of gas}
\label{sec:gas}

The evolution of a barred galaxy containing a
considerable amount of gas in the disc can differ substantially from
that of a galaxy with no gaseous component.

One of the differences is that the pattern speed may stay constant,
or even increase with time, contrarily to what was seen in the purely
stellar cases (Friedli \& Benz 1993, Heller \& Shlosman 1994,
Berentzen et al. 1998). Unfortunately not much quantitative information from
appropriate $N$-body simulations can be found in the 
literature, and a study of the evolution of the pattern speed for
different gas masses and distributions would certainly be most useful.
The problem is particularly interesting since in the exchange of
energy and angular momentum, discussed in the previous section, a third
partner has been added to the stellar disc and the halo components,
namely the gas.

As a response to the bar forcing, the gas concentrates in two very narrow
and elongated regions near the leading edges of the bar, which are in
fact the loci of shocks. Depending on the bar parameters, and in
particular its strength, they are either straight, or curved with
their concave side towards the bar major axis (Athanassoula
1992). As a result of these shocks, the gas flows inwards and
accumulates in an area near the center. The size of this central area is of the
order of a kpc, and is determined by the extent of the largest orbits of
the $x_2$ family. 

In order to come yet further inwards, to the nucleus of the
galaxy, the gas has to shed a yet larger fraction of its angular
momentum. Several possibilities have been so far put forward. In the first
one (Heller \& Shlosman 1994) when the central disc becomes gas
dominated it is unstable and 
breaks into clumps. Via their collisions, these clumps may loose
angular momentum and spiral to the center. A second possibility is
that the $x_2$ family does not exist, or that it
has a very small extent, in which case the central area is
very small. It has, however, been argued that the density distribution
in barred galaxies is such as to permit an $x_2$ family of sizeable
extent 
(e.g. Teuben et al. 1986; Athanassoula 1991, 1992; Athanassoula \&
Bureau 2000). A similar effect can be achieved in the case of a very
high sound speed, in which case the shock loci are very near the
center (Englmeier \& Gerhard 1997, Patsis \& Athanassoula 2000). At
somewhat smaller, but still quite high, sound speeds the distance
between the main shock loci and the center may be bridged by a spiral,
which can also transfer gas to the center (Maciejewski et al. 2002). As
a final alternative let me mention nested bars, which can be either
gaseous, as initially suggested by Shlosman, Frank \&
Begelman (1989), or stellar (Erwin, these proceedings). Recent hydrodynamical 
simulations, however, argue that
secondary stellar bars are unlikely to increase the mass inflow rate
into the galactic nucleus (Heller \& Shlosman 2002, Maciejewski et al.
2002).  Further work on this subject is 
necessary in order to elucidate the properties and the role of secondary
bars.

\section{Bar destruction and regeneration}
\label{sec:destroy}

There are at least two ways by which bars could be destroyed, or, by
which their amplitude could be very severely diminished. One can be 
assimilated to a self-destruction, since the culprit is the 
central concentration formed by the gas which is pushed towards the 
center of the galaxy by the bar itself. In the second method the 
culprit is a companion or small infalling galaxy. 

As mentioned in the previous section, the gas is pushed inwards by the bar.
Once it has reached the center, it may create a sufficient central
concentration to destroy the bar that drove it there (Friedli \& Benz
1993, Berentzen et al. 1998). Indeed, orbital calculations by Hasan \&
Norman (1990) and by Hasan, Pfenniger \& Norman (1993) have shown that
a sufficiently strong central concentration can make the $x_1$ orbits
unstable. Norman, Sellwood \& Hasan (1996) showed that a massive core,
of mass 5\% or more of the combined disc and bulge mass, can destroy a
bar. Nevertheless, the mass of black holes in disc systems seems to be
smaller, by an order of magnitude or more, than what is required by
Norman et al. and thus may not be sufficient to destroy bars.  

In purely stellar simulations, vertical impacts by a massive companion
in the bar area result in a considerable decrease of the bar
amplitude, but destruction was so far found only in cases in which the
disc thickened so considerably that the galaxy could not be called a
disc galaxy anymore (Athanassoula 1996, 1999). On the other hand,
destruction seems to be possible for such impacts if a considerable
gaseous component is present (Berentzen, 
Athanassoula, Heller et al., in preparation). If the companion is
initially in a near-circular orbit, then it can easily destroy the bar
(Athanassoula 1996, 1999), even in simulations with no gaseous component.

Can a disc form a new bar, once its first one has been destroyed? In
other words, can more than one bar form consecutively in the same
disc? It should be noted that after the demise of the first bar, a disc
is inhospitable for a new bar, since it is left with a
considerable central concentration and a high velocity dispersion,
both factors inhibiting bar growth. Nevertheless, fresh infalling gas
may cool the disc and make it again unstable. A first attempt at
making a second bar in this way 
(Sellwood \& Moore 1999) proved successful, so further
work on this subject should follow.

\acknowledgments

I would like to thank A. Bosma, A. Misiriotis, M. Tagger and
F. Masset for stimulating discussions and A. Misiriotis and
J. C. Lambert for their 
collaboration on the software calculating the orbital frequencies. 
Part of this paper was written while I 
was visiting INAOE. I would like to thank the ECOS-Nord and the ANUIES for
financing this trip and INAOE for their kind hospitality.


\begin{references}
\reference Athanassoula, E. 1991, in Dynamics of Disk Galaxies,
  ed. B. Sundelius (G\"oteborg: Chalmers Univ. Technology), 149
\reference Athanassoula, E. 1992, \mnras, 259, 345
\reference Athanassoula, E. 1996, in ASP Conf. Ser. Vol. 91, Barred
  Galaxies, eds. R. Buta, D. Crocker and B. Elmegreen, (San Francisco:
  ASP), 309
\reference Athanassoula, E. 1999, in ASP Conf. Ser. Vol. 160, Astrophysical discs,
  eds. J. A. Sellwood and J. Goodman, (San Francisco: ASP), 160, 351
\reference Athanassoula, E. 2002, \apj, 569, L83 (A02)
\reference Athanassoula, E., Bosma, A., \& Papaioannou, S. 1987, \aap,
   179, 40
\reference Athanassoula, E., \& Bureau, M. 2000, \apj, 522, 699
\reference Athanassoula, E., \& Misiriotis, A. 2002, \mnras, 330, 35 (AM)
\reference Athanassoula, E., \& Sellwood, J. A. 1986, \mnras, 221, 213
\reference Berentzen, I., Heller, C. H., Shlosman, I., \& Fricke
  K. J. 1998, \mnras, 300, 49
\reference Bosma, A. 1999, in ASP conference series 182, Galaxy Dynamics, 
eds. D. R. Merritt, M. Valluri \& J. A. Sellwood, 339
\reference Bosma, A. 2000, in ASP conference Series 197,
>From the Early Universe to the Present, eds. F. Combes,
G. A. Mamon \& V. Charmandaris, (San Francisco: ASP), 91
\reference Bottema, R. 1993, \aap, 275, 16
\reference Combes, F., Debbasch, F., Friedli, D., \& Pfenniger, D. 1990, \aap,
  233, 82
\reference Courteau, S. ,\& Rix, H. W. 1999, \apj, 513, 561
\reference Debattista, V. P., \& Sellwood, J. A. 1998, \apj, 493, L5
\reference Debattista, V. P., \& Sellwood, J. A. 2000, \apj, 543, 704
\reference Englmaier, P., \& Gerhard, O. 1997, \mnras, 287, 57
%\reference Eskridge P. B., Frogel J. A., Pogge R. W. et al 2000, \aj,
%  119, 536
\reference Friedli, D., \& Benz W. 1993, \aap, 268, 65
\reference Hasan, H., \& Norman, C. 1990, \apj, 361, 69
\reference Hasan, H., Pfenniger, D., \& Norman, C. 1993, \apj, 409, 91
\reference Heller, C. H., \& Shlosman, I. 1994, \apj, 424, 84
\reference Heller, C. H., \& Shlosman, I. 2002, \apj, 565, 921
\reference Hernquist, L., \& Weinberg. M. D. 1992, \apj, 400, 80
\reference Hohl, F. 1971, \apj, 168, 343
\reference Kawai, A., Fukushige, T., Makino, J., \& Taiji, M. 2000,
  \pasj, 52, 659  
\reference Kranz, T., Slyz, A. ,\& Rix, H. W. 2001, \apj, 562, 164
\reference Little, B., \& Carlberg, R. G. 1991, \mnras, 250, 161
\reference Lynden-Bell, D., \& Kalnajs, A. J. 1972, \mnras, 157, 1
\reference Maciejewski, W., Teuben, P. J., Sparke L. S., \& Stone,
  J. M. 2002, \mnras, 329, 502
\reference Miller, R. H., Prendergast, K. H., \& Quirk, W. J. 1970,
  \apj 161, 903 
\reference Ohta, K. 1996, in ASP conference Series Vol 91, Barred
Galaxies, eds. R. Buta, D. A. Crocker, B. G. Elmegreen, (San Francisco: ASP), 37
%\reference Ostriker, J. P., \& Peebles, P. J. E. 1973, \apj, 186, 467
\reference Patsis, P., \& Athanassoula, E. 2000, \aap, 358, 45
\reference Sackett, P. D. 1997, \apj, 483, 103
\reference Sellwood, J. A. 1999, in ASP conference series 182, 
Galaxy Dynamics, eds. D. R. Merritt, M. Valluri \& J. A. Sellwood,
(San Francisco: ASP), 351  
\reference Sellwood, J. A., \&  Moore E. M. 1999, \apj, 510, 125
\reference Shlosman, I., Frank, J., \& Begelman, M. C. 1989, 
Nature, 338, 45
\reference Teuben, P., Sanders, R. H., Atherton, P. D., \& van Albada,
  G. D. 1986, \mnras, 221, 1
\reference Tremaine, S., \& Weinberg, M. D. 1984, \mnras, 209, 729
\reference Weinberg, M. D. 1985, \mnras, 213, 451
\reference Weiner, B. J., Sellwood, J. A., \& Williams, T. B. 2001, 
\apj, 546, 931
\reference 
\reference 
\end{references}
\end{document}